\documentclass[3p,times,11pt]{elsarticle}

\usepackage{ecrc}

\volume{00}

\firstpage{1}

\journalname{}
\journal{Statistics and Probability Letters}

\runauth{D. Durante et al.}

\usepackage{amssymb}
 \usepackage{amsthm,amsmath}
\usepackage{setspace}

\usepackage{tikz}
\usetikzlibrary{fit,positioning,calc}
\tikzset{
  font={\fontsize{11pt}{12}\selectfont}}
\usepackage[ruled,norelsize,vlined]{algorithm2e}

\usepackage[colorlinks,citecolor=blue,urlcolor=blue]{hyperref}

\usepackage{graphicx}

\newtheorem{prop}{Proposition}[section]

\def\T{{ \mathrm{\scriptscriptstyle T} }}

\newcommand{\I}{{1}}

\begin{document}

\begin{frontmatter}

\title{{\vspace{-8em} \\A nested  expectation--maximization algorithm \\ for latent class models with covariates}}

\author[a]{Daniele Durante\corref{mycorrespondingauthor}}
\cortext[mycorrespondingauthor]{Corresponding author}
\ead{daniele.durante@unibocconi.it}

\author[b]{Antonio Canale}
\ead{canale@stat.unipd.it}
\author[a]{Tommaso Rigon}
\ead{tommaso.rigon@phd.unibocconi.it}
\address[a]{Department of Decision Sciences, Bocconi University, Via Roentgen 1, 20136 Milan, Italy}
\address[b]{Department of Statistical Sciences, University of Padova, Via Cesare Battisti 241, 35121 Padua, Italy \vspace{-20pt}}

\begin{abstract}
We develop a nested \textsc{em} routine for latent class models with covariates which allows maximization of the full--model log-likelihood and, differently from current methods, guarantees monotone log-likelihood sequences along with improved convergence rates.
\end{abstract}

\end{frontmatter}

\vspace{15pt}
\section{Introduction} 
\label{sec:intro}
Multivariate categorical data are routinely collected in several  fields \citep[e.g.][]{hagen2002}. In these settings, it is of key interest to characterize the dependence structures in the observed data, and to identify underlying classes or subpopulations which may explain these patterns of dependence and their changes with external covariates.  Let  ${\bf Y}_i=(Y_{i1},\ldots, Y_{iJ})^\intercal$ denote the multivariate categorical  random variable generating the  observed data ${\bf y}_i=(y_{i1}, \ldots, y_{iJ})^\intercal \in\ \mathcal{Y} =\{1, \ldots, K_1\}$ $\times \cdots \times \{1, \ldots, K_J\}$, for every unit $i=1, \ldots, n$. Latent class models with covariates \citep[][]{bandeen:etal, day1988, form1992} address this goal by assuming the response variables $Y_{i1},\ldots, Y_{iJ}$, are conditionally independent given a latent class indicator $s_i \in \mathcal{S} =\{1, \ldots, R\}$, whose probability mass function is allowed to change with the covariates ${\bf x}_i=(x_{i1}, \ldots, x_{iP})^{\intercal}$ $\in \mathcal{X}$, under a multinomial logistic regression. Consistent with this assumption, the conditional probability mass function $\mbox{pr}({\bf Y}_i={\bf y} \mid {\bf x}_i)=\mbox{pr}(Y_{i1}=y_{1},\ldots, Y_{iJ}=y_{J} \mid {\bf x}_i)$ for the multivariate random variable ${\bf Y}_i$, can be expressed as 
\begin{eqnarray}
{\small{\mbox{pr}({\bf Y}_i={\bf y} \mid {\bf x}_i) = \sum_{r=1}^R \nu_r({\bf x}_i) \prod_{j=1}^J \pi_{jr}(y_j) = \sum_{r=1}^R \frac{\exp({\bf x}_i^{\intercal} \boldsymbol{\beta}_r)}{\sum_{l=1}^R\exp({\bf x}_i^{\intercal} \boldsymbol{\beta}_l)} \prod_{j=1}^J \pi_{jr}(y_j), \quad \mbox{for each } i=1, \ldots, n,}}
 \label{eq1}
 \end{eqnarray}
 \noindent where  $\nu_r({\bf x}_i)=\mbox{pr}(s_i=r \mid {\bf x}_i) \in (0,1)$ is the covariate--dependent probability of class $r$, whereas $\pi_{jr}(y_j)=\mbox{pr}(Y_{ij}=y_j \mid s_i=r)\in (0,1)$ characterizes the probability to observe the category $y_j$ for the variable $Y_{ij}$ in class $r$. Note also that, consistent with classical multinomial logistic regression, the coefficients vector $ \boldsymbol{\beta}_R=(\beta_{1R}, \ldots, \beta_{PR})^{\intercal}$ associated with the last class $R$ is fixed to zero, in order to avoid identifiability issues. With this choice, ${\bf x}_i^{\intercal}  \boldsymbol{\beta}_r$ measures the log-odds of belonging to class $r$ instead of class $R$, when the vector of covariates is ${\bf x}_i$. Equation \eqref{eq1} provides an interpretable factorization which allows inference on the class-specific generative mechanisms underlying the observed data ${\bf y}_i$ and how the latent classes $s_i$ relate to the covariates ${\bf x}_i$.  Refer to \citep[][]{bandeen:etal, day1988, form1992} for a discussion about this class of models, and to \citep{clog1995} for a review on latent class analysis, including early formulations without covariates \citep{laz1968,mccutcheon1987}.

To obtain the above information, it is necessary to estimate   $\boldsymbol{\beta}=\{\boldsymbol{\beta}_1, \ldots, \boldsymbol{\beta}_{R-1} \}$ and $\boldsymbol{\pi}=\{\pi_{j1}(y_j), \ldots, \pi_{jR}(y_j): j=1, \ldots, J; \ y_j=1, \ldots, K_j \}$. This can be  accomplished by maximizing the log-likelihood function 
\begin{eqnarray}
{\small{
\ell(\boldsymbol{\beta},\boldsymbol{\pi}; {\bf y},{\bf x})= \sum_{i=1}^n\log\left[\sum_{r=1}^R \nu_r({\bf x}_i) \prod_{j{=}1}^J \prod_{y_j{=}1}^{K_j}\pi_{jr}(y_{j})^{\I(y_{ij} = y_j)}\right]= \sum_{i=1}^n\log\left[\sum_{r=1}^R \frac{\exp({\bf x}_i^{\intercal} \boldsymbol{\beta}_r)}{\sum_{l=1}^R\exp({\bf x}_i^{\intercal} \boldsymbol{\beta}_l)}  \prod_{j{=}1}^J \prod_{y_j{=}1}^{K_j}\pi_{jr}(y_{j})^{\I(y_{ij} = y_j)}\right],}}
\label{eq2}
\end{eqnarray}
where $\I(y_{ij}=y_j)$ is $1$ if $y_{ij}=y_j$, and $0$ otherwise. However,  maximization of \eqref{eq2} is not straightforward due to the sum inside the logarithm.  In fact, although some contributions attempt direct maximization of \eqref{eq2} via Newton--Raphson \citep[e.g.][]{haber1988} or  simplex algorithms  \citep[e.g.][]{day1988}, more popular implementations \citep{bandeen:etal,form1992,van1996,bolck2004,vermunt2010} rely on \textsc{em} routines \citep{demp1977}. These strategies leverage a hierarchical specification---equivalent to \eqref{eq1}---which introduces a latent class variable $s_i \in \mathcal{S} =\{1, \ldots, R\}$ for each $i=1, \ldots, n$, to obtain 
\vspace{-12pt}
\begin{eqnarray}
(Y_{ij}  \mid s_i=r) \sim \mbox{\textsc{categorical}}(\boldsymbol{\pi}_{jr}, K_j), \  \mbox{for any } j=1, \ldots, J, \quad \quad (s_i \mid {\bf x}_i)\sim \mbox{\textsc{categorical}}(\boldsymbol{\nu}({\bf x}_i),R),
\label{eq2.2}
\end{eqnarray}
\vspace{-28pt}

\noindent independently for every unit $i=1, \ldots, n$. In \eqref{eq2.2}, $\mbox{\textsc{categorical}}(\boldsymbol{\rho}, H)$ denotes the generic categorical random variable having probability mass function $\boldsymbol{\rho}=\{{\rho}_1, \ldots, {\rho}_H\}$ for the $H$ different categories. Hence, consistent with  \eqref{eq2.2}, if ${\bf s}=(s_1, \ldots, s_n)$ is known, the maximum likelihood estimates for $\boldsymbol{\beta}$ and $\boldsymbol{\pi}$ can be easily obtained by maximizing separately the log-likelihood $\ell_1(\boldsymbol{\beta};  {\bf s},{\bf x})$ associated with the multinomial logistic regression for $s_1, \ldots,s_n$, and the log-likelihood $\ell_2(\boldsymbol{\pi}; {\bf y}, {\bf s})$  for the  categorical data $y_{1j}, \ldots, y_{nj}$, $j=1, \ldots, J$, within each subpopulation defined by the classes in ${\bf s}$. In fact, $\ell(\boldsymbol{\beta},\boldsymbol{\pi};{\bf y},{\bf s}, {\bf x})=\ell_1(\boldsymbol{\beta};  {\bf s},{\bf x})+\ell_2(\boldsymbol{\pi};  {\bf y},{\bf s})$, with
\vspace{-5pt}
\begin{eqnarray}
{\small{
\begin{split}
\ell_1(\boldsymbol{\beta};  {\bf s},{\bf x})+\ell_2(\boldsymbol{\pi};  {\bf y},{\bf s})=\sum_{i=1}^n \sum_{r=1}^R \I(s_i=r)  \log \left[ \frac{\exp({\bf x}_i^{\intercal} \boldsymbol{\beta}_r)}{\sum_{l=1}^R\exp({\bf x}_i^{\intercal} \boldsymbol{\beta}_l)}\right]+ {\sum_{i=1}^n \sum_{r=1}^R}  \left[ {\sum_{j=1}^J \sum_{y_j=1}^{K_j}\I(s_i=r)\I(y_{ij}=y_j) \log  \pi_{jr}(y_{j})} \right].
\end{split}}}
\label{eq3}
\end{eqnarray}

Maximizing $\ell_1(\boldsymbol{\beta};  {\bf s},{\bf x})$ with respect to $\boldsymbol{\beta}$ requires algorithms for multinomial logistic regression, within a generalized linear models framework \citep[e.g.][Chapter 7]{agresti2002}, whereas $\ell_2(\boldsymbol{\pi};  {\bf y},{\bf s})$  is analytically maximized at 
\begin{equation}
\hat{\pi}_{jr}(y_{j})=\frac{\sum_{i=1}^n \I(s_i=r)\I(y_{ij}=y_j)}{{\sum_{i=1}^n} \I(s_i=r),} \quad \mbox{for each } r=1, \ldots, R, \ j=1, \ldots, J, \  y_j=1, \ldots, K_j.
\label{eq:hatpi_1}
\end{equation}
Unfortunately, ${\bf s}$ is not observed. Therefore, estimation of $\boldsymbol{\beta}$ and $\boldsymbol{\pi}$ needs to rely only on the information provided by the data $({\bf y}_i, {\bf x}_{i})$, for $i=1, \ldots, n$.

There are two main strategies in the literature to estimate $\boldsymbol{\beta}$ and $\boldsymbol{\pi}$, generally referred to as one--step \citep{bandeen:etal,form1992,van1996} and three--step \citep{bolck2004,vermunt2010} methods. The former attempt simultaneous estimation of $\boldsymbol{\beta}$ and $\boldsymbol{\pi}$ in  \eqref{eq2}, leveraging the augmented log-likelihood  \eqref{eq3}. The latter consider, instead, a multi--step routine which first estimates  $s_1,  \ldots, s_n$, along with  $\boldsymbol{\pi}$, from a latent class model without covariates, and then uses the predicted classes as responses in $\ell_1(\boldsymbol{\beta};  \hat{{\bf s}},{\bf x})$---or a modification of it---to estimate $\boldsymbol{\beta}$. 

Although the above methods are considered in routine implementations---including the \textsc{R} library \texttt{poLCA} \citep{linzer2011} and the software \textsc{Latent GOLD} \citep{vermunt2016}---as discussed in Sect.~\ref{sec:intro1}--\ref{sec:intro2}, both strategies still raise concerns on the quality of the estimates and on the efficiency of the algorithms. Motivated by these issues, Sect.~\ref{sec:model} describes a nested \textsc{em} which provides more reliable estimation routines for this class of models within the maximum likelihood framework. As outlined on a real dataset in Sect.~\ref{perf}, the proposed methods enjoy improved theoretical properties and superior performance. Concluding remarks can be found in Sect.~\ref{discussion}.

\subsection{One--step estimation methods}
\label{sec:intro1}
Recalling Sect.~\ref{sec:intro}, maximum likelihood estimation for the parameters in the full model  \eqref{eq1}---characterizing one--step methods \citep{bandeen:etal,form1992,van1996}---proceeds via an \textsc{em} algorithm which leverages the complete log-likelihood $\ell(\boldsymbol{\beta},\boldsymbol{\pi};{\bf y},{\bf s}, {\bf x})=\ell_1(\boldsymbol{\beta};  {\bf s},{\bf x})+\ell_2(\boldsymbol{\pi};  {\bf y},{\bf s})$ \eqref{eq3} for the data $({\bf y}_i, {\bf x}_i)$ and the augmented latent class variable $s_i$, $i=1, \ldots, n$. This additive structure of $\ell(\boldsymbol{\beta},\boldsymbol{\pi};{\bf y},{\bf s}, {\bf x})$ allows separate estimation for the parameters $ \boldsymbol{\beta}$ and $\boldsymbol{\pi}$.  Moreover,  $\ell_1( \boldsymbol{\beta}; {\bf s},{\bf x})$ and $\ell_2(\boldsymbol{\pi}; {\bf y},{\bf s})$ are linear in the augmented data $\I(s_{i}=r)$.  Letting $\boldsymbol{\theta}=(\boldsymbol{\beta},\boldsymbol{\pi})$, this facilitates a simple  expectation step in which, at the general iteration $t$, each $\I(s_{i}=r)$  is replaced with
\begin{eqnarray}
{\small{\bar{s}^{(t)}_{ir}=\mbox{E}\{\I(s_{i}=r) \mid  \boldsymbol{\theta}^{(t)}, {\bf y}_i,{\bf x}_i \}=\frac{\exp\{{\bf x}_i^{\intercal} \boldsymbol{\beta}^{(t)}_r\}\prod_{j=1}^J \prod_{y_j=1}^{K_j}\pi^{(t)}_{jr}(y_{j})^{\I(y_{ij}=y_j)}}{\sum_{l=1}^R\exp\{{\bf x}_i^{\intercal} \boldsymbol{\beta}^{(t)}_l\}\prod_{j=1}^J \prod_{y_j=1}^{K_j}\pi^{(t)}_{jl}(y_{j})^{\I(y_{ij}=y_j)}},\quad \mbox{for each } r=1, \ldots, R, \ i=1, \ldots, n,}}
\label{eq:bars}
\end{eqnarray}
to obtain the expected values $Q_1(\boldsymbol{\beta}\mid  \boldsymbol{\theta}^{(t)})$ and $Q_2(\boldsymbol{\pi}\mid  \boldsymbol{\theta}^{(t)})$ of the log-likelihoods $\ell_1(\boldsymbol{\beta};  {\bf s},{\bf x})$ and $\ell_2(\boldsymbol{\pi};  {\bf y},{\bf s})$, respectively, whose summation defines the expectation $Q( \boldsymbol{\beta},\boldsymbol{\pi}\mid  \boldsymbol{\theta}^{(t)})$ of the complete log-likelihood \eqref{eq3}  with respect to the distribution of ${\bf s}$, given the current estimates $\boldsymbol{\theta}^{(t)}$.  Hence, $ \boldsymbol{\theta}^{(t+1)}=\mbox{argmax}_{\boldsymbol{\beta},\boldsymbol{\pi}}\{Q(\boldsymbol{\beta},\boldsymbol{\pi}\mid  \boldsymbol{\theta}^{(t)})\}$ in the \textsc{m}--step can be obtained by maximizing $Q_1(\boldsymbol{\beta}\mid  \boldsymbol{\theta}^{(t)})$ and $Q_2(\boldsymbol{\pi} \mid  \boldsymbol{\theta}^{(t)})$ separately.

Consistent with \eqref{eq:hatpi_1}, the expected log-likelihood $Q_2(\boldsymbol{\pi} \mid \boldsymbol{\theta}^{(t)})$  is easily maximized at
\begin{equation}
{\small{\pi^{(t+1)}_{jr}(y_{j}) = \frac{\sum_{i=1}^n \bar{s}^{(t)}_{ir}\I(y_{ij}=y_j)}{\sum_{i=1}^n \bar{s}^{(t)}_{ir}},  \quad \mbox{for each } r=1, \ldots, R, \ j=1, \ldots, J, \  y_j=1, \ldots, K_j.}}
\label{eq:hatpi}
\end{equation}

It is instead not possibile to maximize analytically $Q_1(\boldsymbol{\beta}\mid \boldsymbol{\theta}^{(t)})$ with respect to $\boldsymbol{\beta}$, due to the logistic link. To address this issue \citep{form1992} and \citep{van1996} consider one Newton--Raphson step relying on a quadratic approximation of $Q_1(\boldsymbol{\beta}\mid \boldsymbol{\theta}^{(t)})$. The routine proposed by   \citep{bandeen:etal} leverages instead a slightly different update relying on the  Hessian and the gradient computed from full--model log-likelihood  in \eqref{eq2}, and evaluated at $(\boldsymbol{\beta}^{(t)},\boldsymbol{\pi}^{(t)})$ and $\bar{s}^{(t)}_{ir}$, $r=1, \ldots, R$, $i=1, \ldots, n$. This procedure---implemented in  \textsc{R} library \texttt{poLCA} \citep{linzer2011}---breaks the \textsc{em} rationale, since the  \textsc{m}--step for $\boldsymbol{\beta}$ maximizes a plug--in estimate of a quadratic approximation for $\ell(\boldsymbol{\beta}; {\bf y},{\bf x})$, instead of $Q_1(\boldsymbol{\beta}\mid \boldsymbol{\theta}^{(t)})$. This may lead to less stable behaviors.

Although the above solutions provide standard approaches to obtain $\boldsymbol{\beta}^{(t+1)}$, the resulting computational routines guarantee neither $Q_1(\boldsymbol{\beta}^{(t+1)}\mid  \boldsymbol{\theta}^{(t)}) \geq Q_1(\boldsymbol{\beta} \mid \boldsymbol{\theta}^{(t)})$, nor  $Q_1(\boldsymbol{\beta}^{(t+1)}\mid \boldsymbol{\theta}^{(t)}) \geq Q_1(\boldsymbol{\beta}^{(t)} \mid \boldsymbol{\theta}^{(t)})$  \citep{bohning1988,bohning1992}, and therefore the proposed methods are neither an \textsc{em}, nor a generalized \textsc{em} algorithm, respectively \citep{demp1977}. Failure to improve the expected log-likelihood may also affect the monotonicity  of the sequence $\ell(\boldsymbol{\beta}^{(t)},\boldsymbol{\pi}^{(t)}; {\bf y},{\bf x})$,  thereby providing  routines which do not meet the basic properties of \textsc{em}, and may not guarantee reliable convergence \citep[e.g.][Chapter 1.5.5]{mclachlan2007}. As we will outline in Sect.~\ref{perf}, this issue is not just found in pathological scenarios, but arises also in routine applications, and may substantially affect convergence to the maximum log-likelihood. Although careful routines  can be designed to overcome this issue, we shall emphasize that  $Q_1(\boldsymbol{\beta}\mid \boldsymbol{\theta}^{(t)})$ is defined on a set of latent responses whose expectation \eqref{eq:bars} changes at every iteration of the algorithm. This setting is more problematic than classical multinomial logit with observed response data, and is further complicated by the need to estimate $\boldsymbol{\pi}$ along with $\boldsymbol{\beta}$ in \eqref{eq1}. Hence, unstable updating steps may have major effects on estimation. Even Monte Carlo \textsc{em} routines \citep{wei1990} do not address the problem, since such methods are devised for intractable \textsc{e}--steps and not for \textsc{m}--steps having no analytical solutions.

To mitigate the above issues, standard implementations consider multiple runs  based on different initializations of the \textsc{em}, and rely on the routine converging to the highest log-likelihood. Alternatively, internal checks can be included to control for decays. These strategies provide more reliable routines, but require multiple runs which increase computational costs. A possibility to reduce decays in $\ell(\boldsymbol{\beta}^{(t)},\boldsymbol{\pi}^{(t)}; {\bf y},{\bf x})$, without relying on multiple runs, is to rescale the inverse of the Hessian by a step--size $0<\alpha\leq 1$ \citep[e.g.][Chapter 1.5.6]{mclachlan2007}.  However, the implementation requires the choice of $\alpha$, without theory on optimal settings. 

Motivated by the above issues, \citep{bohning1988,bohning1992} proposed a Minorize--Majorize (\textsc{mm}) algorithm \citep[e.g.][]{hunt2004} for logistic and multinomial logit regression, which replaces the Hessian with a matrix $\boldsymbol{B}$ to obtain a quadratic function  minorizing the log-likelihood at every $\boldsymbol{\beta}$ and tangent to it in $\boldsymbol{\beta}^{(t)}$. This provides a simple updating scheme similar to the Newton--Raphson which guarantees monotone log-likelihood sequences. Although not currently implemented in latent class models, these strategies can be easily incorporated in the \textsc{m}--step for $\boldsymbol{\beta}$ by replacing the observed responses with the expectation \eqref{eq:bars} of the latent classes.  As outlined in Sect.\ref{perf}, this procedure  provides monotone log-likelihood sequences, but requires more iterations than nested \textsc{em}.

\subsection{Three--step estimation methods}
\label{sec:intro2}
Differently from one--step methods, the three--step procedures \citep[e.g.][]{clog1995} do not attempt direct maximization of \eqref{eq2}, but instead rely on a multi--step strategy which first estimates $\boldsymbol{\pi}$ and the class probabilities via closed--form \textsc{em} for  latent class models without covariates, and then obtain $\hat{\boldsymbol{\beta}}$ from a multinomial logit with the predicted classes $\hat{s}_1, \ldots, \hat{s}_n$ from the previous model acting as observed responses.

The above procedure provides a simple estimation strategy, which is motivated by its interpretability. However, as discussed in \citep{bolck2004} and \citep{vermunt2010}, the resulting estimators for the coefficients in $\boldsymbol{\beta}$ are subject to systematic bias. Motivated by these issues,  \citep{bolck2004} and  \citep{vermunt2010}, recently developed two bias--correction methods relying on a modification of the multinomial log-likelihood for $\hat{s}_1, \ldots, \hat{s}_n$, which incorporates the classification error. The bias--corrected log-likelihood proposed in \citep{bolck2004}  allows simple estimation via standard algorithms for multinomial logistic regression, but this method can be applied only  when the covariates are categorical. A more general correction procedure is considered by \citep{vermunt2010} to include continuous covariates. This approach further improves efficiency and facilitates wider applicability. There is also a focus on other mechanisms to predict the classes, but the final results are not substantially different \citep{vermunt2010, bak2013}.

Although the bias adjustment in \citep{vermunt2010} is widely considered in several implementations,  the estimates  remain still sub-optimal compared to one--step methods, since they do not directly maximize the full--model log-likelihood \eqref{eq2}, and hence cannot be considered as the maximum likelihood estimates. Indeed, when the focus is on providing reliable inference for the parameters $\boldsymbol{\beta}$ and $\boldsymbol{\pi}$ in \eqref{eq1}, it is arguably more coherent to attempt a direct maximization of the log-likelihood in \eqref{eq2}, since it guarantees unbiased, efficient and consistent estimators, under the classical likelihood inference theory, if the model is correctly specified \citep{bolck2004}.

\section{Nested \textsc{em} for one--step estimation} 
\label{sec:model}
To address the aforementioned issues, we propose a nested \textsc{em} algorithm for one--step estimation which avoids approximations of the expected  log-likelihood $Q_1(\boldsymbol{\beta} \mid \boldsymbol{\theta}^{(t)})$, but  improves this function sequentially via a set of  conditional expectation--maximizations for every vector of coefficients $\boldsymbol{\beta}_r$, given the others.

Working with conditional expected log-likelihoods  is appealing in providing a set of different logistic regressions for which the recent P\'olya-gamma data augmentation  scheme \citep{polson2013,choi2013} guarantees closed--form maximization via generalized least squares. Indeed, following Theorem 1 in \citep{polson2013}, the generic logistic likelihood $\exp({\bf x}^{\intercal} \boldsymbol{\beta})^{a}\{1+\exp({\bf x}^{\intercal} \boldsymbol{\beta})\}^{-b}$, can be rewritten as $2^{-b} \exp\{(a-0.5b){\bf x}^{\intercal} \boldsymbol{\beta}\} \mbox{cosh}(0.5{\bf x}^{\intercal} \boldsymbol{\beta})^{-b}$, $b \geq 0$, whereas the likelihood  of a P\'olya-gamma  variable $\omega \sim \textsc{pg}(b,{\bf x}^{\intercal} \boldsymbol{\beta})$ is proportional to $\exp\{-0.5\omega({\bf x}^{\intercal} \boldsymbol{\beta})^2\} \mbox{cosh}(0.5{\bf x}^{\intercal} \boldsymbol{\beta})^{b}$. Hence, combining these two quantities  provides an augmented likelihood $\exp\{-0.5\omega({\bf x}^{\intercal} \boldsymbol{\beta})^2+(a-0.5b){\bf x}^{\intercal} \boldsymbol{\beta}\}$ for data $(a,\omega)$ which is proportional to the one induced by a Gaussian regression for the transformed response $\omega^{-1}(a-0.5b) \sim \mbox{N}({\bf x}^{\intercal} \boldsymbol{\beta}, \omega^{-1})$. This allows simple maximization for $\boldsymbol{\beta}$ within a Gaussian framework. Note that, as discussed in \citep{polson2013}, other Gaussian--related data augmentations for logistic regression have been developed. However, these strategies  require more complex representations \citep{polson2013}, thus leading to intractable computations.  Instead, the P\'olya-gamma  data augmentation leads directly to a Gaussian likelihood with a single latent variable $\omega$, whose expectation is analytically available via $\mbox{E}(\omega)=0.5b({\bf x}^{\intercal} \boldsymbol{\beta})^{-1} \mbox{tanh}(0.5{\bf x}^{\intercal} \boldsymbol{\beta})$. Besides providing tractable computations, the proposed procedure guarantees  monotone log-likelihood sequences and is directly motivated by an exact \textsc{em} for  the special case of $R=2$ classes, described below.

\subsection{Exact \textsc{em} algorithm for $R=2$}
\label{sec:model1}
Let us first focus on deriving  an \textsc{em} for $R=2$ latent classes, which provides analytical maximization also for $\boldsymbol{\beta}_1$. In fact, when $R=2$ the expected value of $\ell_1(\boldsymbol{\beta};  {\bf s},{\bf x})=\ell_1(\boldsymbol{\beta}_1;  {\bf s},{\bf x})$ is
\begin{eqnarray*}
{\small{Q_1(\boldsymbol{\beta} \mid \boldsymbol{\theta}^{(t)})=Q_1(\boldsymbol{\beta}_1 \mid \boldsymbol{\theta}^{(t)})={\sum_{i=1}^n}   \log \left[ \frac{\exp({\bf x}_i^{\intercal} \boldsymbol{\beta}_1)^{\bar{s}^{(t)}_{i1}}}{\{1{+}\exp({\bf x}_i^{\T} \boldsymbol{\beta}_1)\}^{\bar{s}^{(t)}_{i1}}} \cdot \frac{\{1{+}\exp({\bf x}_i^{\intercal} \boldsymbol{\beta}_1)\}^{\bar{s}^{(t)}_{i1}}}{1{+}\exp({\bf x}_i^{\intercal} \boldsymbol{\beta}_1)}\right]=\sum_{i=1}^n \log \left[  \frac{ \exp({\bf x}_i^{\intercal} \boldsymbol{\beta}_1)^{\bar{s}_{i1}^{(t)}}}{1{+}\exp({\bf x}_i^{\intercal} \boldsymbol{\beta}_1)}\right],}}
\end{eqnarray*}
thus providing the log-likelihood $\ell^{*}_1(\boldsymbol{\beta}_1; \bar{\bf s}^{(t)}, {\bf x})$  of a logistic regression in which the entries in $\bar{\bf s}^{(t)}$ act as responses. This allows the implementation of the  P\'olya-gamma data augmentation.  In particular, defining $b:=1$, $a:=\bar{s}_{i1}^{(t)}$, ${\bf x}^{\intercal} \boldsymbol{\beta}:={\bf x}_i^{\intercal} \boldsymbol{\beta}_1$ and $\omega:=\omega_{i1}$, leads to the  complete log-likelihood
{\small{\begin{eqnarray}
\ell^{*}_1(\boldsymbol{\beta}_1; \bar{\bf s}^{(t)}, {\bf x}, \boldsymbol{\omega})&=& \sum_{i=1}^n \log\left[\frac{\mbox{cosh}\{0.5({\bf x}_i^{\intercal} \boldsymbol{\beta}_1)\}}{\exp\{0.5 \omega_{i1}({\bf x}_i^{\T} \boldsymbol{\beta}_1)^2\}}\cdot \frac{\exp\{(\bar{s}^{(t)}_{i1}-0.5){\bf x}_i^{\T} \boldsymbol{\beta}_1\}}{\mbox{cosh}\{0.5({\bf x}_i^{\intercal} \boldsymbol{\beta}_1)\}}\right] + \mbox{const}, \nonumber\\
&=& \sum_{i=1}^n[- 0.5\omega_{i1}({\bf x}_i^{\intercal} \boldsymbol{\beta}_1 )^2+(\bar{s}^{(t)}_{i1}-0.5){\bf x}_i^{\intercal} \boldsymbol{\beta}_1]+ \mbox{const},
\label{eq:l1j2}
\end{eqnarray}}}
\noindent for $\bar{\bf s}^{(t)}$, ${\bf x}$, and the P\'olya-gamma augmented data $\boldsymbol{\omega}=(\omega_{11}, \ldots, \omega_{n1})$. Equation \eqref{eq:l1j2} is a quadratic function of ${\bf x}_i^{\intercal} \boldsymbol{\beta}_1$, and is linear in $\boldsymbol{\omega}$. This allows the implementation of a simple nested expectation step in which every $\omega_{i1}$ is replaced with the expectation  $\mbox{E}(\omega_{i1} \mid  \boldsymbol{\pi}^{(t)}, \boldsymbol{\beta}_1^{(t)}, {\bf y}_i,{\bf x}_i ):=\bar{\omega}^{(t)}_{i1}
=0.5 ({\bf x}_i^{\intercal} \boldsymbol{\beta}^{(t)}_1)^{-1}\mbox{tanh}(0.5{\bf x}_i^{\intercal} \boldsymbol{\beta}^{(t)}_1) $  to obtain $Q^*_1(\boldsymbol{\beta}_1 \mid \boldsymbol{\theta}^{(t)}) = \sum_{i=1}^n-0.5\bar{\omega}^{(t)}_{i1}(\bar{\eta}^{(t)}_{i1}-{\bf x}_i^{\intercal} \boldsymbol{\beta}_1)^2 + \mbox{const}$, with $\bar{\eta}^{(t)}_{i1} = (\bar{s}_{i1}^{(t)}-0.5)/\bar{\omega}^{(t)}_{i1}$. The appealing property associated with this nested expected log-likelihood,  compared to  $Q_1(\boldsymbol{\beta}_1\mid \boldsymbol{\theta}^{(t)})$, is that it allows direct maximization for $\boldsymbol{\beta}_1$. Indeed, exploiting the generalized least squares, $Q^*_1(\boldsymbol{\beta}_1 \mid \boldsymbol{\theta}^{(t)})$ is maximized at 

\vspace{-10pt}
\begin{equation}
{\small{\boldsymbol{\beta}_1^{(t+1)}=({\bf X}^{\intercal} \bar{\boldsymbol{\Omega}}^{(t)}{{\bf X}})^{-1}{ \bf X}^{\intercal}\bar{\boldsymbol{\Omega}}^{(t)}\bar{\boldsymbol{\eta}}^{(t)},}}
\label{eq13}
\end{equation}

\vspace{0pt}

\noindent where ${\bf X}$ is the $n \times P$ matrix having rows ${\bf x}^{\intercal}_i$, whereas $\bar{\boldsymbol{\Omega}}^{(t)}=\mbox{diag}(\bar{\omega}^{(t)}_{11}, \ldots, \bar{\omega}^{(t)}_{n1})$ and $\bar{\boldsymbol{\eta}}^{(t)}=(\bar{\eta}^{(t)}_{11}, \ldots, \bar{\eta}^{(t)}_{n1})^{\intercal}$. Refer to \cite{dur2018} for recent results proving that solution \eqref{eq13} guarantees monotone convergence at  optimal rate in logistic regression. This provides additional support for \eqref{eq13}, compared to Newton--Raphson and \textsc{mm} updates.

We shall stress that, although we obtain $Q^*_1(\boldsymbol{\beta}_1 \mid \boldsymbol{\theta}^{(t)})$ sequentially, this quantity coincides with the expectation of the complete log-likelihood $\ell_1(\boldsymbol{\beta}_1; {\bf s}, {\bf x}, \boldsymbol{\omega})$. Hence, since $Q_2(\boldsymbol{\pi} \mid \boldsymbol{\theta}^{(t)})$ is analytically maximized in \eqref{eq:hatpi}, the resulting routine is an exact  \textsc{em} algorithm  based on the complete log-likelihood $\ell(\boldsymbol{\beta}_1,\boldsymbol{\pi}; {\bf y},{\bf s},{\bf x}, \boldsymbol{\omega})=\ell_1(\boldsymbol{\beta}_1; {\bf s}, {\bf x}, \boldsymbol{\omega})+\ell_2(\boldsymbol{\pi}; {\bf y}, {\bf s})$. Finally, note that since $\boldsymbol{\pi}^{(t+1)}$ can be easily obtained before updating $\boldsymbol{\beta}$, a more efficient strategy, inspired by the multi-cycle expectation conditional maximization algorithm \citep{meng1993}, is to update also the expectation of ${\bf s}$ based on $\boldsymbol{\pi}^{(t+1)}$ instead of $\boldsymbol{\pi}^{(t)}$, before applying \eqref{eq13}. As discussed in Sect.~\ref{sec:model2}, this solution will be adopted in the general case with $R>2$. 

\subsection{Nested  \textsc{em} algorithm for $R>2$}
\label{sec:model2}
When more than two classes are considered, $Q_1(\boldsymbol{\beta} \mid \boldsymbol{\theta}^{(t)})$ has a multinomial logit form, and not a logistic one. Thus,  a direct application of the P\'olya-gamma data augmentation is not possible. However, as we will outline, the conditional expected log-likelihood for every vector of coefficients $\boldsymbol{\beta}_r$, given the others,
 can be rewritten as a proper logistic log-likelihood \citep[e.g.][]{holmes2006}, thus motivating a P\'olya-gamma data augmentation. Based on this result,  we propose a nested \textsc{em} which improves the expected log-likelihood for $\boldsymbol{\beta}$ via a set of conditional expectation--maximizations. In particular, for each iteration $t$, we consider $R^{*}=R-1$ nested cycles which sequentially improve the conditional expected log-likelihood of one vector of coefficients $\boldsymbol{\beta}_r$, fixing the other $\boldsymbol{\beta}_l$, $l \neq r$ at their most recent value. Hence, let 
$\boldsymbol{\beta}^{(t+r/R^{*})}=\{\boldsymbol{\beta}_1^{(t+1)}, \ldots, \boldsymbol{\beta}^{(t+1)}_r,\boldsymbol{\beta}^{(t)}_{r+1}, \ldots, \boldsymbol{\beta}^{(t)}_{R^{*}} \},$
denote the estimates for the class-specific vectors of coefficients at cycle $r=1, \ldots, R^{*}$ in iteration $t$, we seek a sequential updating procedure providing the chain inequalities

\vspace{-25pt}

\begin{eqnarray}
{\small{Q_1(\boldsymbol{\beta}^{(t+r/R^{*})}\mid \boldsymbol{\pi}^{(t+1)}, \boldsymbol{\beta}^{\{t+(r-1)/R^{*}\}})  \geq Q_1(\boldsymbol{\beta}^{\{t+(r-1)/R^{*}\}}\mid \boldsymbol{\pi}^{(t+1)}, \boldsymbol{\beta}^{\{t+(r-1)/R^{*}\}}), \quad \mbox{for each } r=1, \ldots, R^{*}.}}
\label{eq:ineq}
\end{eqnarray}

\vspace{-5pt}

\noindent The key difference between $\boldsymbol{\beta}^{(t+r/R^{*})}$ and $\boldsymbol{\beta}^{\{t+(r-1)/R^{*}\}}$ in \eqref{eq:ineq}, is that only the coefficients in $\boldsymbol{\beta}_r$ are updated, whereas all the others are kept fixed. Hence, at every cycle $r$ we seek to improve the expected log-likelihood produced in $r-1$, by modifying only $\boldsymbol{\beta}_r$ from its previous estimate at $t$ to a new one at $t+1$.  In this respect, such strategy partially recalls the  univariate version of Newton--Raphson in \citep{vermunt2016}, and adapts it to blocks of parameters to obtain more efficient updates relying on the most recent estimates. Moreover, as outlined in Proposition~\ref{prop1}, the sequential improvements underlying the nested \textsc{em}, along with  the direct maximization of $Q_2(\boldsymbol{\pi} \mid \boldsymbol{\theta}^{(t)})$, guarantee the monotonicity of the sequence $\ell(\boldsymbol{\beta}^{(t)},\boldsymbol{\pi}^{(t)}; {\bf y},{\bf x})$. This is fundamental to ensure reliable convergence \citep{bohning1988}.  Note that in  \eqref{eq:ineq}, also the expectation of the conditional  log-likelihood with respect to the augmented data is sequentially updated using the estimates of the coefficients from the previous cycle, in the same spirit of the multi-cycle expectation conditional maximization  \citep{meng1993}.

In deriving an updating procedure for $\boldsymbol{\beta}$ with property  \eqref{eq:ineq}, we adapt results in Sect.~\ref{sec:model1}, which provides simple and explicit maximization for each $\boldsymbol{\beta}_r$. Focusing on cycle $r$ within iteration $t$, let $Q_1(\boldsymbol{\beta}_r\mid \boldsymbol{\theta}^{\{t+(r-1)/R^{*}\}})$, with $\boldsymbol{\theta}^{\{t+(r-1)/R^{*}\}}=( \boldsymbol{\pi}^{(t+1)}, \boldsymbol{\beta}^{\{t+(r-1)/R^{*}\}})$, denote the conditional expected log-likelihood, written as a function of only $\boldsymbol{\beta}_r$, with all the other class-specific coefficients fixed at their corresponding estimates at cycle $r-1$. According to equations \eqref{eq3} and \eqref{eq:bars}, the function $Q_1(\boldsymbol{\beta}_r\mid \boldsymbol{\theta}^{\{t+(r-1)/R^{*}\}})$ can be expressed as
\begin{eqnarray}
{\small{\sum_{i=1}^n \left\{ \bar{s}^{\{t+(r-1)/R^{*}\}}_{ir}  \log \left[ \frac{\exp({\bf x}_i^{\intercal} \boldsymbol{\beta}_r)}{\exp({\bf x}_i^{\intercal} \boldsymbol{\beta}_r)+c_i^{\{t+(r-1)/R^{*}\}}}\right] +\sum_{l\neq r} \bar{s}^{\{t+(r-1)/R^{*}\}}_{il}  \log \left[ \frac{\exp({\bf x}_i^{\intercal} \boldsymbol{\beta}^{\{t+(r-1)/R^{*}\}}_l)}{\exp({\bf x}_i^{\intercal} \boldsymbol{\beta}_r)+c_i^{\{t+(r-1)/R^{*}\}}}\right]\right\},}}
\label{q_cond}
\end{eqnarray}
where the constants $c_i^{\{t+(r-1)/R^{*}\}}$ denote the sum of all the exponential quantities $\exp({\bf x}_i^{\intercal} \boldsymbol{\beta}_l)$, $l\neq r$,  written as a function of the current estimates for the class-specific coefficients at cycle $r-1$, whereas the expectation of the augmented latent class indicators are calculated in \eqref{eq:bars} as a function of $\boldsymbol{\pi}^{(t+1)}$ and $\boldsymbol{\beta}^{\{t+(r-1)/R^{*}\}}$. 

Since we aim to improve $Q_1(\boldsymbol{\beta}^{(t)}_r\mid \boldsymbol{\theta}^{\{t+(r-1)/R^{*}\}})=Q_1(\boldsymbol{\beta}^{\{t+(r-1)/R^{*}\}}\mid \boldsymbol{\theta}^{\{t+(r-1)/R^{*}\}})$ by updating the current estimate of $\boldsymbol{\beta}_r$ under the P\'olya-gamma data augmentation outlined in the previous section, let us  highlight a logistic log-likelihood in \eqref{q_cond}. Indeed, holding out  additive constants not depending on $\boldsymbol{\beta}_r$, and dividing both the numerator and the denominator of the arguments in the logarithmic functions by the quantities in the vector ${\bf c}^{\{t+(r-1)/R^{*}\}}$, we easily obtain
\begin{eqnarray}
Q_1(\boldsymbol{\beta}_r\mid \boldsymbol{\theta}^{\{t+(r-1)/R^{*}\}}) =\sum_{i=1}^n \log \left[  \frac{\{ \exp({\bf x}_i^{\intercal} \boldsymbol{\beta}_r- a_i^{\{t+(r-1)/R^{*}\}})\}^{\bar{s}^{\{t+(r-1)/R^{*}\}}_{ir}}  }{1+\exp({\bf x}_i^{\intercal} \boldsymbol{\beta}_r-a_i^{\{t+(r-1)/R^{*}\}})}\right]+ \mbox{const}, 
\label{eq:q1}
\end{eqnarray}
where $a_i^{\{t+(r-1)/R^{*}\}}= \log c_i^{\{t+(r-1)/R^{*}\}},$  and provided that $\bar{s}^{\{t+(r-1)/\bar{R}\}}_{ir}+\sum_{l\neq r} \bar{s}^{\{t+(r-1)/\bar{R}\}}_{il}=1$. Hence, up to the constants $a_i^{\{t+(r-1)/\bar{R}\}}$ in the linear predictor, equation \eqref{eq:q1} for $\boldsymbol{\beta}_r$ has the same form of the expected log-likelihood  for $ \boldsymbol{\beta}_1$ in  Sect.~\ref{sec:model1}, thereby motivating the P\'olya-gamma data augmentation at each cycle $r$ of the nested \textsc{em}. In particular, introducing  $\omega_{ir}\ {\sim}\ \textsc{pg}(1,{\bf x}_i^{\intercal} \boldsymbol{\beta}_r - a_i^{\{t+(r-1)/R^{*}\}})$, for every $i=1, \ldots, n$, we can exploit the analytical results in Sect.~\ref{sec:model1}, to show that the conditional expectation of each $\omega_{ir}$ is
\vspace{-10pt}
\begin{eqnarray}
{{\bar{\omega}^{\{t+(r-1)/R^{*}\}}_{ir}=0.5({\bf x}_i^{\intercal}  \boldsymbol{\beta}^{(t)}_r- a_i^{\{t+(r-1)/R^{*}\}})^{-1}\mbox{tanh}[0.5({\bf x}_i^{\intercal}  \boldsymbol{\beta}^{(t)}_r- a_i^{\{t+(r-1)/R^{*}\}})],}}
\label{eq:baromega}
\end{eqnarray}
\vspace{-25pt}

\noindent and that the desired increment  \eqref{eq:ineq}  at cycle $r$, can be simply obtained---similarly to \eqref{eq13}---by setting
\vspace{-10pt}
\begin{eqnarray}
{\small{\boldsymbol{\beta}_r^{(t+1)}= ({\bf X}^{\intercal} \bar{\boldsymbol{\Omega}}^{\{t+(r-1)/R^{*}\}}{\bf X})^{-1}{\bf X}^{\intercal}\bar{\boldsymbol{\Omega}}^{\{t+(r-1)/R^{*}\}}\bar{\boldsymbol{\eta}}^{\{t+(r-1)/R^{*}\}},}}
\label{eq13b}
\end{eqnarray}
\vspace{-25pt}

\noindent where $\bar{\boldsymbol{\Omega}}^{\{t+(r{-}1)/R^{*}\}}$ is an $n \times n$ diagonal matrix with elements $\bar{\omega}^{\{t+(r-1)/R^{*}\}}_{ir}$, whereas $\bar{\boldsymbol{\eta}}^{\{t+(r-1)/R^{*}\}}$ is a vector of length $n$ with entries $\bar{\eta}^{\{t+(r-1)/R^{*}\}}_{ir}= (\bar{s}^{\{t+(r-1)/R^{*}\}}_{ir}-0.5+\bar{\omega}^{\{t+(r-1)/R^{*}\}}_{ir}a^{\{t+(r-1)/R^{*}\}}_{i})/\bar{\omega}^{\{t+(r-1)/R^{*}\}}_{ir}$, for $i=1, \ldots, n$.

\begin{algorithm*}[t]
\caption{Nested \textsc{em} algorithm for latent class regression models.} \label{al1}
{\small{\begin{tabbing}
   \enspace Initialize the model parameters $\boldsymbol{\pi}^{(1)}$ and $\boldsymbol{\beta}^{(1)}$ at iteration $t=1$. Then, for $t=1$ until convergence of $\ell(\boldsymbol{\beta}^{(t)},\boldsymbol{\pi}^{(t)}; {\bf y},{\bf x}) $\\
   \qquad \textsc{\bf Expectation}: compute $\bar{s}^{(t)}_{ir}$ as in \eqref{eq:bars}.\\
   \qquad  \textsc{\bf Maximization:} update the current estimate $\boldsymbol{\pi}^{(t)}$ to obtain $\boldsymbol{\pi}^{(t+1)}$ as in \eqref{eq:hatpi}. \\
   \qquad For $r=1$ to $r=R^{*}$ \\
   \qquad\qquad  \textsc{\bf Nested Expectation}:  compute $\bar{s}^{\{t+(r-1)/R^{*}\}}_{ir}$ and $\bar{\omega}^{\{t+(r-1)/R^{*}\}}_{ir}$ by applying  \eqref{eq:bars} and \eqref{eq:baromega} to   $\boldsymbol{\pi}^{(t+1)}$   and the \\  
     \qquad\qquad  current  estimates of $\boldsymbol{\beta}$ produced by cycle $r-1$. \\
     \qquad\qquad   \textsc{\bf Nested Maximization:} update the current estimate $\boldsymbol{\beta}_r^{(t)}$ to obtain $\boldsymbol{\beta}_r^{(t+1)}$ as in \eqref{eq13b}.
\end{tabbing}}}
\vspace{-10pt}
\end{algorithm*}

Algorithm \ref{al1} provides details for the implementation of nested \textsc{em}. Note that all the steps require simple and exact expressions, in contrast with current routines. Moreover, as proved in Proposition~\ref{prop1}, our methods imply monotone convergence. 
\begin{prop}
The nested \textsc{em} in  Sect. \ref{sec:model2} implies $\ell(\boldsymbol{\beta}^{(t+1)},\boldsymbol{\pi}^{(t+1)}; {\bf y},{\bf x}) \geq \ell(\boldsymbol{\beta}^{(t)},\boldsymbol{\pi}^{(t)}; {\bf y},{\bf x})$ at any $t$.
\label{prop1}
\end{prop}
\vspace{-15pt}

\begin{proof}[Proof]
To prove Proposition~\ref{prop1}, we start by showing that the nested conditional expectation--maximizations for each $\boldsymbol{\beta}_r$, $r=1, \ldots, R^{*}$, imply  \eqref{eq:ineq}. Since each of these optimizations steps is a pure \textsc{em} based on the P\'olya-gamma data augmentation,  following \citep{demp1977}, we have $\ell^{*}_1(\boldsymbol{\beta}^{(t+1)}_r; \bar{\bf s}^{\{t+(r-1)/R^{*}\}}_{r}, {\bf x}) \geq \ell^{*}_1(\boldsymbol{\beta}^{(t)}_r; \bar{\bf s}^{\{t+(r-1)/R^{*}\}}_{r}, {\bf x})$ for each  $r=1, \ldots, R^{*}$. This implies $Q_1(\boldsymbol{\beta}^{\{t+r/R^{*}\}}{\mid} \boldsymbol{\theta}^{\{t+(r-1)/R^{*}\}}) \geq Q_1(\boldsymbol{\beta}^{\{t+(r-1)/R^{*}\}}\mid \boldsymbol{\theta}^{\{t+(r-1)/R^{*}\}})$, provided that in \eqref{eq:q1} these log-likelihoods differ by an additive constant which does not depend on $\boldsymbol{\beta}_r$. To conclude the proof, we need to ensure that the inequalities in \eqref{eq:ineq}, along with the direct maximization of $Q_2(\boldsymbol{\pi} \mid \boldsymbol{\theta}^{(t)})$, guarantee $\ell(\boldsymbol{\beta}^{(t+1)},\boldsymbol{\pi}^{(t+1)}; {\bf y},{\bf x}) \geq \ell(\boldsymbol{\beta}^{(t)},\boldsymbol{\pi}^{(t)}; {\bf y},{\bf x})$. Let $Q(\boldsymbol{\beta},\boldsymbol{\pi} \mid \boldsymbol{\theta}^{(t)})=Q_1(\boldsymbol{\beta} \mid \boldsymbol{\theta}^{(t)})+Q_2(\boldsymbol{\pi}  \mid \boldsymbol{\theta}^{(t)})$ be the expected log-likelihood, written as a function of all the parameters in our model. Direct maximization of $Q_2(\boldsymbol{\pi} \mid \boldsymbol{\theta}^{(t)})$ in the first step of our routine, guarantees $Q(\boldsymbol{\pi} ^{(t)}, \boldsymbol{\beta}^{(t)} \mid \boldsymbol{\pi} ^{(t)}, \boldsymbol{\beta}^{(t)}) \leq Q(\boldsymbol{\pi} ^{(t+1)}, \boldsymbol{\beta}^{(t)} \mid \boldsymbol{\pi} ^{(t)}, \boldsymbol{\beta}^{(t)})$. Therefore, as discussed in page 165 of \citep{mclachlan2007}, this first result guarantees $\ell(\boldsymbol{\pi} ^{(t+1)}, \boldsymbol{\beta}^{(t)}; {\bf y},{\bf x}) \geq \ell(\boldsymbol{\pi} ^{(t)}, \boldsymbol{\beta}^{(t)}; {\bf y},{\bf x})$. Similarly, the inequalities in  \eqref{eq:ineq}, characterizing each nested cycle $r$, ensure $\ell(\boldsymbol{\pi} ^{(t+1)}, \boldsymbol{\beta}^{(t+r/R^{*})} ; {\bf y},{\bf x}) \geq \ell(\boldsymbol{\pi} ^{(t+1)},\boldsymbol{\beta}^{(t+(r-1)/R^{*})} ; {\bf y},{\bf x})$ for every $r=1, \ldots, R^{*}$. Joining these results we can prove Proposition \ref{prop1} via 
\begin{equation*}
{\small{
\begin{split}
&\ell(\boldsymbol{\pi} ^{(t)}, \boldsymbol{\beta}_1^{(t)},  \ldots,\boldsymbol{\beta}_r^{(t)}, \ldots ,\boldsymbol{\beta}_{R^{*}}^{(t)} ;  {\bf y},{\bf x}) \leq \ell(\boldsymbol{\pi} ^{(t+1)}, \boldsymbol{\beta}_1^{(t)}, \ldots, \boldsymbol{\beta}_r^{(t)}, \ldots ,\boldsymbol{\beta}_{R^{*}}^{(t)} ;  {\bf y},{\bf x}) \leq \ell(\boldsymbol{\pi} ^{(t+1)}, \boldsymbol{\beta}_1^{(t+1)},\ldots, \boldsymbol{\beta}_r^{(t)}, \ldots ,\boldsymbol{\beta}_{R^{*}}^{(t)} ;  {\bf y},{\bf x}) \leq\\
& \ldots \leq \ell(\boldsymbol{\pi} ^{(t+1)}, \boldsymbol{\beta}_1^{(t+1)} ,\ldots, \boldsymbol{\beta}_r^{(t+1)}, \ldots ,\boldsymbol{\beta}_{R^{*}}^{(t)} ;  {\bf y},{\bf x}) \leq \ldots  \leq  \ell(\boldsymbol{\pi} ^{(t+1)}, \boldsymbol{\beta}_1^{(t+1)}, \ldots, \boldsymbol{\beta}_r^{(t+1)}, \ldots ,\boldsymbol{\beta}_{R^{*}}^{(t+1)} ;  {\bf y},{\bf x}).  
\end{split}}}
\end{equation*}
\vspace{-24pt}
\end{proof}

\subsection{Hybrid nested \textsc{em} for one--step estimation}
\label{hybrid}
Before evaluating the empirical performance of nested \textsc{em}, we first propose a more practical and efficient hybrid modification. Indeed,  as discussed in \citep{vermunt2010}, the \textsc{em}  is characterized by stable maximization even when the initialization is far from the optimal solution, whereas Newton--Raphson  guarantees fast convergence when the routine is close to the maximum. Motivated by this result, we propose an hybrid procedure which starts with nested \textsc{em} and then switches to classical Newton--Raphson \citep{form1992,van1996}  when the increment  $\ell(\boldsymbol{\beta}^{(t+1)},\boldsymbol{\pi}^{(t+1)}; {\bf y},{\bf x})-\ell(\boldsymbol{\beta}^{(t)},\boldsymbol{\pi}^{(t)}; {\bf y},{\bf x})$ is less than or equal to a pre--specified small $\epsilon \geq 0$. Although the inclusion of Newton--Raphson steps could  still cause decays in the log-likelihood sequence, when the routine is in a neighborhood of the global---or local---maximum, the parabolic approximation is more stable. 

\section{Empirical study}
\label{perf}
\vspace{-2pt}
To study the benefits of nested \textsc{em} in a real--data application, we compare its computational performance with those of the routines discussed in Sect.~\ref{sec:intro1} and \ref{sec:intro2}. In particular, one--step competitors comprise the \textsc{em} with one Newton--Raphson step (\textsc{nr}EM) from \citep{bandeen:etal}, the more formal  \textsc{em} (\textsc{nr}EM$Q_1$) maximizing $Q_1(\boldsymbol{\beta}\mid  \boldsymbol{\theta}^{(t)})$ as in \citep{form1992,van1996}, and the conservative version of \textsc{nr}EM$Q_1$ which relies on the rescaled updating of $\boldsymbol{\beta}$, as outlined in Sect.~\ref{sec:intro1}, with $\alpha=0.5$ (\textsc{nr}EM$Q_1$, $\alpha=0.5$). Recalling Sect.~\ref{sec:intro1}, we additionally adapt the \textsc{mm} (\textsc{mm}EM) proposed by \citep{bohning1988,bohning1992} to compare empirical performance with a routine guaranteeing monotone log-likelihood sequences. The three--step methods considered are instead the classical strategy (\textsc{3step}Classical) discussed in Sect.~\ref{sec:intro2} \citep[e.g.][]{clog1995}, and the popular bias--corrected algorithm (\textsc{3step}Correction)  from \citep{vermunt2010}. The \textsc{hybrid}EM is instead the modification of our \textsc{nested}EM presented in Sect.~\ref{hybrid}. In performing maximization under this modified version of the nested \textsc{em} we set $\epsilon=0.01$. Although we have found the results robust to moderate changes in small values  of $\epsilon$, it is important to notice that high  $\epsilon$ should be avoided since it allows the hybrid routine to rely on Newton--Raphson steps even when $\ell(\boldsymbol{\beta}^{(t)},\boldsymbol{\pi}^{(t)}; {\bf y},{\bf x})$ is far from the maximum.

To provide a detailed assessment, we perform estimation under the above algorithms for 100 runs, with varying random initialization. The routines are all initialized at the same values---for each run---and stop when the increment in the log-likelihood is lower than $10^{-11}$. For every run  we study the maximization performance  and the computational efficiency of the different algorithms. Specifically, the maximization performance is monitored by the number of runs with a decay in the log-likelihood sequence, and by the frequency of runs converging to local modes. For the runs reaching local modes, we also compute the median of the absolute difference between the  log-likelihood in these local modes and the maximum one. Computational efficiency is instead studied via the median  number of iterations for convergence, computed only for the runs reaching ${\mbox{max}_{{\boldsymbol{\beta, \pi}}}}\{\ell({\boldsymbol{\beta,\pi}}; {\bf y},{\bf x})\}$, and the averaged computational time.\footnote{Computations rely on \textsc{R} (version 3.3.2) implementations in a machine with 1 Intel Core i5 2.5 GHz processor and 4 GB \textsc{ram}.} Code to reproduce the analyses and further results are available at \url{https://github.com/danieledurante/nEM}.

\begin{table*}
\caption{Quantitative evaluation of the maximization quality and the computational efficiency of the routines discussed in Sect.~\ref{sec:intro1}--\ref{sec:model}. The value $\mbox{max}_{{\boldsymbol{\theta}}}\{\ell({\boldsymbol{\theta}}; {\bf y},{\bf x})\}$ coincides with the highest log-likelihood observed in the 100 runs of the different algorithms. Since most of these routines are devised to maximize $\ell({\boldsymbol{\theta}}; {\bf y},{\bf x})$, such choice can be safely regarded as the maximum log-likelihood.}
\label{tab:2}       
{\small{
\begin{tabular}{lllll}
\hline\noalign{\smallskip}
{\bf ELECTION DATA $R=3$}& \textsc{nr}EM & \textsc{nr}EM$Q_1$  & \textsc{nr}EM$Q_1$ $\alpha=0.5$  & \textsc{mm}EM    \\
\noalign{\smallskip}\hline\noalign{\smallskip}
Number of runs with a decay in  $\ell({\boldsymbol{\theta}}^{(t)}; {\bf y},{\bf x})$ & 78 &37  & 12  & 0    \\
Number of runs reaching a local mode & 94  & 51 & 24 & 27  \\
$|\ell(\hat{\boldsymbol{\theta}}; {\bf y},{\bf x}){-}\mbox{max}_{{\boldsymbol{\theta}}}\{\ell({\boldsymbol{\theta}}; {\bf y},{\bf x})\}|$ local modes  & 830.046 & 1105.626 & 4.894  & 0.644 \\
Iterations to reach $\mbox{max}_{{\boldsymbol{\theta}}}\{\ell({\boldsymbol{\theta}}; {\bf y},{\bf x})\}$  & 146 &  151 &  161 & 229   \\
Averaged computational time for each run& 0.073'' &  0.182''&   0.230''&  0.283''  \\
\noalign{\smallskip}\hline
\noalign{\smallskip}
{\bf ELECTION DATA $R=3$}& \textsc{3step}Classical & \textsc{3step}Correction  & \textsc{nested}EM & \textsc{hybrid}EM      \\
\noalign{\smallskip}\hline\noalign{\smallskip}
Number of runs with a decay in  $\ell({\boldsymbol{\theta}}^{(t)}; {\bf y},{\bf x})$ & \textsc{na} &  \textsc{na} & 0  & 0   \\
Number of runs reaching a local mode &100 &100  & 24  & 25   \\
$|\ell(\hat{\boldsymbol{\theta}}; {\bf y},{\bf x}){-}\mbox{max}_{{\boldsymbol{\theta}}}\{\ell({\boldsymbol{\theta}}; {\bf y},{\bf x})\}|$ local modes & 42.224&39.104 & 0.644  &  0.644\\
Iterations to reach $\mbox{max}_{{\boldsymbol{\theta}}}\{\ell({\boldsymbol{\theta}}; {\bf y},{\bf x})\}$    &  \textsc{na} &\textsc{na}   & 171  & 166   \\
Averaged computational time for each run&0.229'' & 0.212'' &0.359''   &  0.265''  \\
\noalign{\smallskip}\hline
\end{tabular}}}
\vspace{-10pt}
\end{table*}

Table \ref{tab:2} summarizes the performance of the methods discussed in Sect.~\ref{sec:intro1}--\ref{sec:model}, for an application to the \textsc{election} data available in the \textsc{R} library \texttt{poLCA} \citep{linzer2011}. This dataset measures voters political affiliation, along with their opinions on how well six different personality traits  describe the candidates Al Gore and George Bush before the 2000 presidential elections. These $J=12$ categorical opinions are collected on a four items scale for $n=880$ voters. Here, we assess performance of the maximization routines considering $R=3$ classes, with the political affiliation entering as covariate in the multinomial logit for such latent classes.

Consistent with Sect.~\ref{sec:intro1}--\ref{sec:model}, \textsc{nested}EM and  \textsc{mm}EM always provide monotone sequences for \eqref{eq2}, thereby guaranteeing accurate maximization and reduced frequency of local modes. Including a Newton--Raphson step as in  \citep{bandeen:etal,form1992,van1996}, leads instead to decays in the log-likelihood sequences, which increase the chance of local modes far from  ${\mbox{max}_{{\boldsymbol{\beta}, \boldsymbol{\pi}}}}\{\ell({\boldsymbol{\beta},\boldsymbol{\pi}}; {\bf y},{\bf x})\}$. As discussed in Sect.~\ref{sec:intro1}, this issue is more severe for  \textsc{nr}EM \citep{bandeen:etal} compared to the formal  \textsc{nr}EM$Q_1$ \citep{form1992,van1996}. This reduced maximization performance of the Newton--type algorithms  is mitigated by partial improvements in computational efficiency compared to \textsc{nested}EM  and \textsc{mm}EM, which remain, however, on a similar scale both for the averaged computational time of each run, and for the number of iterations to reach ${\mbox{max}_{{\boldsymbol{\beta}, \boldsymbol{\pi}}}}\{\ell({\boldsymbol{\beta},\boldsymbol{\pi}}; {\bf y},{\bf x})\}$. Since these algorithms perform estimation in fractions of seconds, an improved maximization performance is arguably the most important property. 

Combining \textsc{nested}EM and \textsc{nr}EM$Q_1$, provides an \textsc{hybrid}EM which guarantees the accurate maximization performance of monotone algorithms and  a computational efficiency comparable to Newton--type methods. Also rescaling the Newton--Raphson updating of $\boldsymbol{\beta}$ by $\alpha=0.5$,  allows improvements in maximization performance, but these gains are associated with a reduced computational efficiency. To  conclude the evaluation of the one--step algorithms, we shall notice that the correction proposed by \citep{bohning1988,bohning1992} guarantees reliable maximization, but, as theoretically proved in \citep{dur2018}, their global conservative bound requires more iterations to reach convergence compared to  \textsc{nested}EM  and  \textsc{hybrid}EM.

As discussed in Sect.~\ref{sec:intro2}, the three--step algorithms do not attempt direct maximization of \eqref{eq2}. The consequences of this are evident in Table~\ref{tab:2} with all the runs of the \textsc{3step}Classical and the \textsc{3step}Correction routines providing sub-optimal estimates which induce local modes in the full--model log-likelihood. Due to this, it is not possible to study the number of iterations to reach ${\mbox{max}_{{\boldsymbol{\beta}, \boldsymbol{\pi}}}}\{\ell({\boldsymbol{\beta},\boldsymbol{\pi}}; {\bf y},{\bf x})\}$. Also the number of decays in  $\ell({\boldsymbol{\beta}}^{(t)},{\boldsymbol{\pi}}^{(t)}; {\bf y},{\bf x})$ is somewhat irrelevant to evaluate the three--step methods, since the estimation routines are based on two separate maximizations not  directly related to   \eqref{eq2}. It is also worth noticing that, although our studies are based on the log-likelihood sequence instead of the parameters estimates, within a maximum likelihood approach these two quantities are directly related. In fact, our goal is not on providing a new class of models and estimators for the parameters in \eqref{eq1}, but an improved maximization routine for likelihood--based inference, with more reliable convergence to the optimal parameters estimates maximizing \eqref{eq2}. In this respect, we have also inspected the estimates  associated with the local modes in Table~\ref{tab:2} observing evident deviations from the optimal ones, especially in local maxima with a log-likelihood notably below  ${\mbox{max}_{{\boldsymbol{\beta}, \boldsymbol{\pi}}}}\{\ell({\boldsymbol{\beta},\boldsymbol{\pi}}; {\bf y},{\bf x})\}$. This result provides further support to our nested \textsc{em}.

In performing the above studies we initialized $\boldsymbol{\beta}$ from independent Gaussians with mean $0$ and small variance $0.5$. Reducing such variance---i.e. initializing $\boldsymbol{\beta}$ close to $0$---yield to gains in all  routines. Also in these cases, however, the performance of  \textsc{nested}EM and  \textsc{hybrid}EM was not worse than the other algorithms, while being less sensitive to  initialization.  Similar conclusions were obtained in other applications.

\section{Discussion}
\label{discussion}

Motivated by the recent P\'olya-gamma data augmentation for logistic regression, and by the lack of fully reliable estimation routines for latent class  models with covariates, we developed a nested \textsc{em} for one--step maximum likelihood estimation within this class of models. Differently from Newton--type methods, the nested \textsc{em} has theoretical guarantee of monotone log-likelihood sequences, while ensuring convergence at a faster rate than \textsc{mm} routines \citep{dur2018}. This provides improved estimation and computational efficiency.

Although our focus has been on latent class analysis with covariates, the nested \textsc{em} can be also adapted to multinomial logit models with Gaussian random effects. In this setting, the calculation of the expected log-likelihood for the fixed coefficients requires intractable marginalization over the  Gaussian random effects, thus requiring Monte Carlo \textsc{em} \citep{wei1990}, h--likelihood  \citep{lee2006} and other methods \citep[e.g.][]{mccul1997}. Our nested \textsc{em} could provide key benefits over these alternatives. In fact, conditioned on P\'olya-gamma augmented data, it is possible to perform closed--form marginalization and maximization as in Gaussian linear mixed models.

{\setstretch{1}
\bibliographystyle{plainnat}

}

\end{document}